\documentclass[12pt]{article}

\usepackage{array,dsfont} 
\usepackage{epsfig}
\usepackage{amssymb}
\usepackage{graphics,graphpap}

\renewcommand{\thefootnote}{\fnsymbol{footnote}}

\begin{document}

\begin{titlepage}
\begin{flushright}\begin{tabular}{l}
IPPP/06/16\\
DCPT/06/32\\
CERN--PH--TH/2006--050
\end{tabular}
\end{flushright}
\vskip1.5cm
\begin{center}
   {\Large \bf \boldmath $|V_{td}/V_{ts}|$ from $B\to V\gamma$}
    \vskip1.3cm {\sc
Patricia Ball$^{1,2}$\footnote{Patricia.Ball@durham.ac.uk} and Roman
Zwicky$^{1}$\footnote{Roman.Zwicky@durham.ac.uk}}
  \vskip0.5cm
{\em        $^1$ IPPP, Department of Physics,
University of Durham, Durham DH1 3LE, UK \\
\vskip0.4cm
$^2$ CERN, CH--1211 Geneva 23, Switzerland}\\
\vskip2.5cm 

\vskip3cm

{\large\bf Abstract\\[10pt]} \parbox[t]{\textwidth}{
The dominant theoretical uncertainty in extracting $|V_{td}/V_{ts}|$
from the ratio of branching ratios $R\equiv{\cal B}(B\to
(\rho,\omega)\gamma)/{\cal B}(B\to
K^* \gamma)$ is given by the ratio of form factors $\xi\equiv T_1^{B\to
  K^*}(0)/T_1^{B\to \rho}(0)$. We re-examine $\xi$ in
the framework of QCD sum rules on the light-cone, taking into account
hitherto neglected SU(3)-breaking effects. We find $\xi 
= 1.17\pm 0.09$. Using QCD factorisation for the branching ratios, 
and the current experimental average for $R$ quoted by HFAG, 
this translates into  
$|V_{td}/V_{ts}|^{\rm HFAG}_{B \to V \gamma} = 0.192\pm 0.014 ({\rm th})
\pm 0.016({\rm exp})$. 
This result agrees, within errors, with that obtained from the Standard Model 
unitarity triangle, $\left| V_{td}/ V_{ts} \right|_{\rm SM}= 0.216\pm  0.029$,
based on tree-level-only processes, and with $|V_{td}/V_{ts}|_{\Delta m} = 
0.2060^{+0.0081}_{-0.0060}({\rm th})\pm 0.0007({\rm exp})$, 
from the CDF measurement of $B_s$ oscillations.  \\[0.5cm]
{\bf This version differs from the original version of the paper, 
published as JHEP 04 (2006) 046, by the inclusion of 
the new BaBar measurement of \boldmath $ 
B \to \rho(\omega) \gamma$ presented at ICHEP 2006, which significantly
shifts the results for $|V_{td}/V_{ts}|$.}
}

\vfill

\end{center}
\end{titlepage}

\setcounter{footnote}{0}
\renewcommand{\thefootnote}{\arabic{footnote}}

\newpage

\section{Introduction}\label{sec:1}

Recently, the Belle collaboration measured the $b\to d$
penguin-dominated decay $B\to(\rho\,,\omega)\gamma$ \cite{Belle},
whereas BaBar obtained an upper bound in 2004 \cite{BaBar} and 
presented a measurement at ICHEP 2006 \cite{BaBar2}. Assuming the Standard
Model (SM) to be valid, this process offers the possibility to extract the
CKM matrix element $|V_{td}|$, in complementarity to the
determination from $B_d$ mixing and the SM unitarity triangle
based on $|V_{ub}/V_{cb}|$ and the angle $\gamma$. 
In order to extract $|V_{td}|$ from
the measured rate, one needs to know both short-distance weak and strong
interaction effects and long-distance QCD effects. Whereas the former
can, at least in principle, 
be calculated to any desired precision in the framework of
effective field theories, and actually  are currently known to
next-to-leading order in
QCD \cite{SD}, the assessment of long-distance QCD effects is
notoriously difficult. After a long history of phenomenologically or
$1/N_c$-motivated factorisation formulas, QCD factorisation
\cite{BVga,BoBu} has provided a consistent framework allowing one to
write the relevant hadronic matrix elements as
\begin{equation}\label{1}
\langle V\gamma|Q_i| B\rangle =
\left[ T_1^{B\to V}(0)\, T^I_{i} +
\int^1_0 d\xi\, du\, T^{II}_i(\xi,u)\, \phi_B(\xi)\, \phi_{V;\perp}(v)\right]
\cdot\epsilon\,.
\end{equation}
Here $\epsilon$ is the photon polarisation 4-vector, $Q_i$ is one of
the operators in the  effective Hamiltonian,
$T_1^{B\to V}$ is a $B\to V$ transition form factor,
and $\phi_B$, $\phi_{V;\perp}$ 
are leading-twist light-cone distribution amplitudes
of the $B$ meson and the vector meson $V$, respectively.
These quantities are universal non-perturbative objects and
describe the long-distance dynamics of the matrix elements, which
is factorised from the perturbative short-distance interactions
included in the hard-scattering kernels $T^I_{i}$ and $T^{II}_i$.
The above QCD factorisation formula is valid in the heavy-quark limit
$m_b\to\infty$ and is subject to 
corrections of order $\Lambda_{\rm QCD}/m_b$. Although it is possible
to determine $|V_{td}|$ from the branching ratio of
$B\to(\rho,\omega)\gamma$ itself, the associated theoretical
uncertainties get greatly reduced when one considers the ratio of
branching ratios for $B\to K^*\gamma$ and $B\to (\rho,\omega)\gamma$
instead. One then can extract $|V_{td}/V_{ts}|$ from
\begin{equation}\label{Brat}
\frac{{\cal B}(B\to(\rho,\omega)\gamma)}{{\cal B}(B\to K^*\gamma)} =
\left|\frac{V_{td}}{V_{ts}}\right|^2
\left(\frac{1-m_{\rho,\omega}^2/m_B^2}{1-m_{K^*}^2/m_B^2}\right)^3
\left( \frac{T_1^{\rho,\omega}(0)}{T_1^{K^*}(0)}\right)^2 \left [ 1 +
  \Delta R\right],
\end{equation}
where the estimates of $\Delta R$ available in the literature lie
between, approximately, 0 and 0.2 \cite{BVga,BoBu}. $\Delta R$ 
contains all non-factorisable effects
induced by $T_i^{I,II}$ in (\ref{1}). As $|V_{ts}|=|V_{cb}|$ in the
SM, up to a small correction $\sim 2\%$, and
$|V_{cb}|$ is known with a precision of 2\% \cite{Vcb}, 
$|V_{td}|$ follows immediately from $|V_{td}/V_{ts}|$. 
The theoretical uncertainty of
this determination is governed by both the ratio of form factors
$T_1^{K^*}(0)/T_1^{\rho,\omega}(0)$ and the value of $\Delta R$, which
parametrises not only SU(3)-breaking effects, but also power-suppressed
corrections to the QCD factorisation formula. The aim of the present
paper is to re-examine the size of SU(3)-breaking corrections to
$T_1^{K^*}(0)/T_1^{\rho,\omega}(0)$ from QCD sum rules on the
light-cone and to determine a value of $|V_{td}/V_{ts}|$ from
(\ref{Brat}), evaluating $\Delta R$ in QCD factorisation; 
we will address the issue of
power-suppressed corrections to $\Delta R$ 
in a separate publication \cite{prep2}.

Our paper is organised as follows: in Section~\ref{sec:2} we discuss
the QCD sum rule for the ratio of form factors
$T_1^{K^*}(0)/T_1^{\rho,\omega}(0)$ and its dependence on
SU(3)-breaking parameters. In Section~\ref{sec:3} we extract a value
of $|V_{td}/V_{ts}|$ from the experimental branching ratio, using QCD
factorisation for the calculation of $\Delta R$. We summarise in
Section~\ref{sec:4}. The appendix contains some formulas  relevant to
the calculation of NLO evolution of twist-2 vector-meson light-cone 
distribution amplitudes.

\section{\boldmath The Form-factor Ratio $T_1^{B\to K^*}/T_1^{B\to
    \rho}$}\label{sec:2}

In this section, we present a concise formula for the form factor
$T_1^{B\to V}(0)$, as obtained from QCD sum rules on the light-cone, and
discuss the hadronic quantities that enter this expression. We do not
discuss the technique of QCD sum rules itself, or that of QCD sum
rules on the light-cone, for which we refer to the literature
\cite{reviews}. Suffice it to say  that the light-cone sum rule for $T_1$
is based on the light-cone expansion of the 
correlation function of the chromomagnetic dipole 
operator $Q_7$ and the interpolating field $\bar q
i\gamma_5 b$ of the $B$ meson. The expansion is in terms of the
convolution of 
process-specific perturbative
kernels  and universal meson light-cone distribution amplitudes (DAs)
of the final-state vector meson, which are ordered in terms of
increasing twist. These DAs have been studied in
Refs.~\cite{BBKT,BB98_2}, mostly for the $\rho$ meson, including two-
and three-particle Fock states up to twist 4. An extension to the
$K^*$ meson is in preparation \cite{prep}. The light-cone expansion is
matched to the description of the correlation function in terms of hadrons
by analytic continuation
into the physical regime and the application of a Borel
transformation, introducing the Borel parameter $M^2$ and
exponentially suppressing contributions from higher-mass states.
In order to extract the contribution
of the $B$ meson, one describes that of other hadron states by
a continuum model, which introduces a second model parameter, 
the continuum threshold $s_0$. The sum rule then yields 
the form factor in question, $T_1$, multiplied by the coupling of the
 $B$ meson to
its interpolating field, i.e.\ the $B$ meson's leptonic decay constant $f_B$.  
At tree level, the sum rule for $T_1^{B\to V}(0)$ then
reads, to twist-4 accuracy:
\begin{eqnarray}
\lefteqn{\frac{m_B^2 f_B}{m_b}\, T_1^{B\to V}(0) e^{-m_B^2/M^2} = 
f_V^\perp m_b \int_{u_0}^1 du e^{-m_b^2/(uM^2)}
\,\frac{\phi_\perp(u)}{2u}} \nonumber\\
&&{}+ f_V^\parallel m_V \int_{u_0}^1 du e^{-m_b^2/(uM^2)}\left[
  \frac{\Phi(u)}{2u} + \frac{1}{2}\,g_\perp^{(v)}(u) +
  \frac{1}{8u}\left( 1 - u \frac{d}{du}\right)
  g_\perp^{(a)}(u)\right.\nonumber\\
&&\hspace*{1.3cm}{}
\left. -\frac{1}{u}\frac{d}{du} \int_0^u d\alpha_1 \int_0^{\bar
    u}d\alpha_2 \,\frac{u-\alpha_1}{2\alpha_3^2}\,
\left({\cal A}(\underline{\alpha})\vphantom{\frac{1}{2}} +
  {\cal V}(\underline{\alpha})\right)\right]\nonumber\\
&&{}+ f_V^\perp m_b \frac{m_V^2}{m_b^2}  \int_{u_0}^1 du
e^{-m_b^2/(uM^2)}\left[\frac{1}{2}\frac{d}{du}\left\{
  \vphantom{\frac{1}{2}} u\bar u
  \phi_\perp(u) + 2 I_L(u) + u H_3(u)
  \right.\right.\nonumber\\
&&\hspace*{1.3cm}\left.{}-\int_0^u d\alpha_1 \int_0^{\bar u}
  d\alpha_2 \,\frac{1}{\alpha_3} \left(
  S(\underline{\alpha}) - \tilde S(\underline{\alpha}) +
  T_1^{(4)}(\underline{\alpha}) - T_2^{(4)}(\underline{\alpha}) + 
T_3^{(4)}(\underline{\alpha}) - 
T_4^{(4)}(\underline{\alpha})\right)\right\}\nonumber\\
&&{}\hspace*{1.3cm}\left. -
  \frac{1}{8}\, u \frac{d^2}{du^2}\,{\mathbb A}_\perp(u)\right],\\
&\equiv & m_b \int_{u_0}^1 du e^{-m_b^2/(uM^2)}\left[ f_V^\perp  R_1(u)
  + f_V^\parallel\,\frac{m_V}{m_b} \,R_2(u) + f_V^\perp
  \left(\frac{m_V}{m_b}\right)^2 R_3(u)\right],\label{SR}
\end{eqnarray}
where $u_0$ is given by $m_b^2/s_0$. $f_V^\parallel$ 
and $f_V^\perp$ are the decay constants of, respectively,
longitudinally and transversely polarised vector mesons.
$\phi_\perp$, $\Phi$, $g_\perp^{(v,a)}$,
$I_L$ and $H_3$ are two-particle distribution amplitudes and integrals
thereof, as defined in Ref.~\cite{BZ04}. 
$\cal A$, $\cal V$, $S$, $\tilde S$ and $T_i^{(4)}$
are three-particle DAs. $u$ is the longitudinal momentum fraction of
the quark in a two-particle Fock state of the final-state vector
meson, whereas $\alpha_{1,2,3}$, with $\sum \alpha_i=1$, are the
momentum fractions of the partons in a three-particle state. The
light-cone expansion is accurate up to terms of order $(m_V/m_b)^3$. Although
we only write down the tree-level expression for the form factor, radiative
corrections are known for $R_1$ \cite{BB98} and the two-particle
contributions to $R_2$ \cite{BZ04}, and will be included in the
numerical analysis. All scale-dependent quantities are calculated at
the (infra-red) 
factorisation scale $\mu^2_F = m_B^2-m_b^2$. The form factor itself
carries an ultra-violet scale dependence, which however cancels in the
ratio.

It is clearly visible from the above formula that the respective
weight of various contributions is controlled by the parameter
$m_V/m_b$; the next term in the light-cone expansion contains 
twist-3, -4 and -5 DAs and is 
of order $(m_V/m_b)^3$. Nonetheless, (\ref{SR}) cannot be
interpreted as $1/m_b$ expansion: for $m_b\to\infty$, the support
of the integrals in $u$ also becomes of ${\cal O}(1/m_b)$, as $1-u_0\sim
1 - m_b^2/s_0 \sim \omega_0/m_b$, with $\omega_0\sim 1\,$GeV 
a hadronic quantity \cite{emili}. 
In this case, the scaling of the various terms in
$m_b$ is controlled by the behaviour of the DAs near the end-point
$u\to 1$. For finite $m_b$, however, the sum rules are not sensitive
to the details of the end-point behaviour, as we shall see
below. Numerically, the expansion in terms of $m_V/m_b$ works very
well and is a reformulation of the ordering of contributions in terms
of the parameter $\delta$ introduced in Ref.~\cite{BZ04}.

We have already discussed $T_1$ in Refs.~\cite{BB98,BZ04};
in the present paper we focus on the ratio
\begin{equation}\label{xi}
\xi \equiv \frac{T_1^{B\to K^*}(0)}{T_1^{B\to \rho}(0)}\,,
\end{equation}
which governs the extraction of $|V_{ts}/V_{td}|$ from $B\to V\gamma$
decays.
Our sum rules can of course be used to determine each form factor
separately, but we expect the ratio to be more accurate, because 
$\xi$ is independent of the $B$-meson decay constant $f_B$
and also, to very good accuracy, of $m_b$ and the sum rule parameters 
$M^2$ and $s_0$; we shall come back to that point below.
Hence, in this paper, we will not re-analyse the absolute
values of $T_1^{B\to (\rho,K^*)}(0)$
nor, consequently, the branching ratios themselves. However, $\xi$ is
very sensitive to SU(3)-breaking effects in the DAs, and 
it is precisely these effects we shall focus on in this paper. A
similar analysis for the ratio of the $D\to K$ and $D\to\pi$ form
factors was carried
out in Ref.~\cite{0608}.

Compared with our previous results of Refs.~\cite{BB98,BZ04}, 
in this paper we implement the 
following improvements:
\begin{itemize}
\item updated values of SU(3)-breaking in twist-2 parameters;
\item complete account of SU(3)-breaking in twist-3 and -4 DAs;
\item estimate of higher-order conformal contributions to twist-4 DAs,
  using the renormalon model of Ref.~\cite{renormalon};
\item NLO evolution for twist-2 parameters.
\end{itemize}
Before presenting numerical results for $\xi$,
let us first discuss the values of the hadronic input parameters
collected in Table~\ref{tab:1}. First of all, we would like to 
mention that we will not distinguish between
the form factors of $\rho$ and $\omega$. 
Their difference is mainly caused by different values of the decay
constants, $f^{\parallel(\perp)}_\rho\neq
f^{\parallel(\perp)}_\omega$, whose precise determination, e.g\ from
experimental data for $\omega\to e^+ e^-$, is complicated by 
mixing with the $\phi$ meson.
In the present paper we take the view that the uncertainty 
induced by letting $T_1^{B\to\rho} = T_1^{B\to\omega}$ 
is negligible compared to 
other uncertainties.

\begin{table}[tbp]
\renewcommand{\arraystretch}{1.2}\addtolength{\arraycolsep}{0pt}
$$
\begin{array}{l|l|l|l||l|l|l||l|l}
   & \rho & \mu = 1\,{\rm GeV} & {\rm Ref.} & K^* & \mu = 1\,{\rm
  GeV} & {\rm Ref.} & {\rm Order} & {\rm Remarks}\\\hline\hline
R_1 & f_\rho^\perp & 0.165\pm0.009 & \mbox{TP}
& f_{K^*}^\perp & 0.185\pm 0.010
  & \cite{BZa1} & \mbox{twist-2} & \mbox{in units of GeV}\\
& a_1^\perp(\rho) & 0 & & a_1^\perp(K^*) & 0.04\pm 0.03 & \cite{BZa1} & 
  \mbox{twist-2} & \mbox{G-odd}\\
& a_2^\perp(\rho) & 0.15\pm 0.07 & \mbox{TP} & a_2^\perp(K^*) &
  0.11\pm 0.09 & \mbox{TP} & \mbox{twist-2} &
  a_2^\perp(K^*)-a_2^\perp(\rho)\\
&&&&&&&& \mbox{constrained}\\
& \Delta_\rho^\perp & 1.24\pm 0.11 & \mbox{TP} & 
\Delta_{K^*}^\perp & 1.18\pm 0.14 & \mbox{TP}
& \mbox{twist-2} & \mbox{BT model  \cite{angi}}\\
& p^\perp_\rho & 3 & & p^\perp_{K^*} & 3 &
   & \mbox{twist-2} & \mbox{for $\phi_\perp$}\\
\hline
R_2 & f_\rho^\parallel & 0.209\pm 0.002
& {\rm exp.} & f_{K^*}^\parallel & 0.217\pm
  0.005 & {\rm exp.} & \mbox{twist-2} &
  \mbox{in units of GeV}\\
& a_1^\parallel(\rho) & 0 & & a_1^\parallel(K^*) & 0.03\pm 0.02 &
  \cite{BZa1} &  \mbox{twist-2} & \mbox{G-odd}\\
& a_2^\parallel(\rho) & = a_2^\perp(\rho) & \mbox{TP} & a_2^\parallel(K^*) &
  = a_2^\perp(K^*) & \mbox{TP} & \mbox{twist-2} &
  a_2^\parallel(K^*)-a_2^\parallel(\rho)\\
&&&&&&&& \mbox{constrained}\\
& \Delta_\rho^\parallel & =\Delta_\rho^\perp & 
& \Delta_{K^*}^\parallel &
  =\Delta_{K^*}^\perp & & \mbox{twist-2} &\mbox{BT model \cite{angi}}\\
& p^\parallel_\rho & =p^\perp_\rho &  & p^\parallel_{K^*} &
  =p^\perp_{K^*} & & \mbox{twist-2} &\mbox{for
  $\phi_\parallel$} \\
& \zeta^A_{3\rho} & 0.032\pm 0.010 & \cite{ZZC} & \zeta^A_{3K^*} & 
(1.0\pm 0.1)\zeta^A_{3\rho} & \mbox{TP} & \mbox{LO twist-3} & \mbox{UR}\\
& \kappa_{3\rho}^\parallel & 0 & &  \kappa_{3K^*}^\parallel & 0.001\pm
  0.001 & \mbox{TP} & \mbox{LO twist-3} & \mbox{G-odd, UR}\\
& \omega^A_{3\rho} & -2\pm 2 & \cite{ZZC} & 
\omega^A_{3K^*} & =\omega^A_{3\rho} & & 
\mbox{NLO twist-3} &  \mbox{UR}\\
 & \omega^V_{3\rho} & \phantom{-}4\pm 2 
& \cite{ZZC} &  \omega^V_{3K^*} & =\omega^V_{3\rho} & & 
\mbox{NLO twist-3} &  \mbox{UR}\\
 & \lambda^A_{3\rho} & 0 & &  \lambda^A_{3K^*} & 0\pm 2 & \mbox{TP} & 
\mbox{NLO twist-3} &  \mbox{G-odd, UR}\\
 & \lambda^V_{3\rho} & 0 & &  \lambda^V_{3K^*} & 0\pm 2 & \mbox{TP} & 
\mbox{NLO twist-3} &  \mbox{G-odd, UR}\\\hline
R_3 & \kappa_{3\rho}^\perp & 0 & & \kappa_{3K^*}^\perp & 0\pm 0.01
  &  \mbox{TP} & \mbox{LO twist-3} & \mbox{G-odd, UR}\\
& \omega^T_{3\rho} & 7\pm 7 & \cite{BBKT} &  
\omega^T_{3K^*} & =\omega^T_{3\rho} & & 
\mbox{NLO twist-3} &  \mbox{UR}\\
 & \lambda^T_{3\rho} & 0 & &  \lambda^T_{3K^*} & 0\pm 2 & \mbox{TP} &
  \mbox{NLO twist-3} &  \mbox{G-odd, UR}\\
& \zeta^T_{4\rho} & 0.10\pm 0.05 & \cite{BBK} 
& \zeta^T_{4K^*} & =\zeta^T_{4\rho}
& & \mbox{LO twist-4} & \mbox{UR}
\end{array}
$$
\caption[]{Hadronic parameters entering $R_{1,2,3}$ in the sum rule
  for $T_1$, Eq.~(\ref{SR}). For twist-2 DAs, we use both the conformal
  expansion Eq.~(\ref{eq:confexp}), truncated after $n=2$, and the
  model of Ball and Talbot (BT) \cite{angi} given
  in terms of two parameters, $\Delta$ and $p$.
  The values of $a_2(\rho)$ and $a_2(K^*)$ are highly correlated; we
  fix $a_2(K^*)-a_2(\rho)=-0.04\pm 0.02$ for both longitudinal and
  transverse DAs. The twist-3 and -4 G-odd parameters
  have never been considered before; all twist-3 and -4
  parameters are under revision (UR) and will be considered in full
  detail in Ref.~\cite{prep}. In this paper (TP), twist-3 SU(3)-breaking 
  effects are only taken into account at LO in the conformal expansion.
  Higher-orders in the conformal expansion of twist-4
  DAs are calculated in the renormalon model of
  Ref.~\cite{renormalon}, see text.}\label{tab:1}
\end{table}

The longitudinal decay constants $f_{\rho,K^*}^\parallel$ 
can be extracted from
the experimental decay rates 
$\tau^-\to (\rho^-,K^{*-}) \nu_\tau$ as \cite{PDG}
$$f_\rho^\parallel = (0.209\pm 0.002)\,{\rm GeV},\qquad
 f_{K^*}^\parallel = (0.217\pm
0.005)\, {\rm GeV}.$$
There is no direct experimental measurement of the tensor decay
constants $f_{\rho,K^*}^\perp$, 
which instead have to be determined from non-perturbative
methods such as lattice calculations \cite{fperplatt1,fperplatt2} 
or QCD sum rules
\cite{elena,BZa1}. Lattice results are available in the quenched
approximation with a chirally improved lattice Dirac operator, which
allows one to reach small quark masses, and for the ratio of decay constants
$f_V^\perp/f_V^\parallel$ \cite{fperplatt2}; 
a first study for the $\rho$ with dynamical fermions
was reported in Ref.~\cite{unquenched}. 
One result of these calculations is that the
ratio of decay constants
only weakly depends on the quark masses. For the $\rho$,
Ref.~\cite{fperplatt2} quotes
$$
\left( \frac{f_{\rho}^\perp}{f_\rho^\parallel}\right)_{\rm latt} 
(2\,{\rm GeV}) = 0.72\pm 0.02\,,
$$
obtained for the lattice spacing $a=0.15\,$fm. 
As for QCD sum rules, the value
$f_\rho^\perp(1\,{\rm GeV}) = (0.160\pm 0.010)\,{\rm GeV}$
was obtained in Ref.~\cite{BB96}. For the present paper, we have re-analysed
the corresponding sum rules, using updated values of $\alpha_s$ and
NLO evolution of $f_\rho^\perp$, and find
\begin{equation}\label{x1}
f_\rho^\perp(1\,{\rm GeV}) = (0.165\pm 0.009)\,{\rm GeV}.
\end{equation}
Also $f_\rho^\parallel$ can be calculated from sum rules, yielding
$(0.206\pm 0.007)\,$GeV. If one calculates the ratio directly from QCD
sum rules, one finds\footnote{The NLO scaling factor
  $f^\perp(2\,{\rm GeV})/f^\perp(1\,{\rm GeV})$ is $0.876$.}
$$
\left( \frac{f_{\rho}^\perp}{f_\rho^\parallel}\right)_{\rm SR} 
(2\,{\rm GeV}) = 0.69\pm
0.04\,,
$$
in agreement with the lattice result.

The determination of $f_{K^*}^\perp$
is less straightforward, see Ref.~\cite{BZa1}, where
\begin{equation}\label{x2}
f_{K^*}^\perp(1\,{\rm GeV}) = (0.185\pm 0.010)\,{\rm GeV}
\end{equation}
was obtained. Evaluating the ratio $f_{K^*}^\perp/f_{K^*}^\parallel$
directly from sum rules, we find
$$
\left( \frac{f_{K^*}^\perp}{f_{K^*}^\parallel}\right)_{\rm SR} 
(2\,{\rm GeV}) = 0.73\pm 0.04\,,
$$
which agrees with the interpolation between the corresponding results
for $\rho$ and $\phi$ obtained from lattice \cite{fperplatt2}. 

Summarising, it is probably fair to say that the present status of
$f_V^\perp$ decay constants is not entirely satisfactory. The
accuracy of the QCD sum rule estimates is unlikely to improve, so any
significant reduction of uncertainty has to come from
lattice. For the moment, however, all existing lattice results still come
with considerable uncertainty (no continuum limit, no results for
$K^*$ with chirally improved Dirac operator), so that in the numerical
analysis of $\xi$ we will use
the experimental results for $f_{\rho,K^*}^\parallel$
and the QCD sum rule results (\ref{x1}) and (\ref{x2}) for
$f_{\rho,K^*}^\perp$.

As for twist-2 DAs, the standard approach is to parametrise them in
terms of a few parameters which are the leading-order terms in
the conformal expansion
\begin{equation}\label{eq:confexp}
\phi(u,\mu^2) = 6 u (1-u) \left( 1 + \sum\limits_{n=1}^\infty
  a_{n}(\mu^2) C_{n}^{3/2}(2u-1)\right).
\end{equation}
To leading-logarithmic accuracy the (non-perturbative)
Gegenbauer moments $a_n$ renormalize multiplicatively. This feature is
due to the conformal symmetry of massless QCD at one-loop, 
the $a_n$ start to mix at next-to-leading order, see
appendix. Although (\ref{eq:confexp}) is not an expansion in any
obvious small parameter, the contribution of terms with large $n$ to
physical amplitudes is suppressed by the fact that the Gegenbauer
polynomials  oscillate rapidly and hence are ``washed out'' upon
integration over $u$ with a ``smooth'' (i.e.\ not too singular) 
perturbative hard-scattering
kernel. For vector mesons, one usually takes into account the terms
with $n=1,2$; the $a_n$ are estimated from QCD sum rules which are
known to become less reliable for larger $n$. As an alternative, one
can build models for $\phi$ based on an assumed fall-off behaviour of
$a_n$ for large $n$.  The model of Ball and Talbot (BT)
\cite{angi}, for instance, 
assumes that, at a certain reference
scale, e.g.\ $\mu=1\,$GeV, the $a_n$ fall off
as powers of $n$:
$$
a_{2n} \propto \frac{1}{(n + 1)^p}.
$$
BT fix the 
absolute normalisation of the Gegenbauer moments  by the first
inverse moment:
$$\int_0^1 \frac{du}{2u}\,\left(\phi(u)+\phi(1-u)\right) 
\equiv 3 \Delta = 3 \left(1 + \sum_{n=1}^\infty a_{2n}\right),$$
which can be viewed as a convolution with the singular hard-scattering 
kernel $1/u$ and gives all $a_n$ the same (maximum) weight $1$.
The rationale of this model is that the DA is given in terms of only
two parameters, $p$ and $\Delta$, and allows one to estimate the
effect of higher order terms in the conformal expansion on
observables. In this paper, we calculate the form factor using both
conformal expansion, truncated after $n=2$, and the BT model, 
normalised by $a_2$ and taking into account terms up to $n=8$. We
shall see below that the effect of terms with $n>2$ is very small.

For the $\rho$, $a_2^{\perp,\parallel}$ have been determined in
Ref.~\cite{BB96}. In the present study we have re-examined the corresponding
sum rules and find, at the scale $\mu=1\,$GeV,
$a_2^\perp(\rho) = 0.15\pm 0.07$ and  $a_2^\parallel(\rho) = 0.14\pm 0.06$,
which is slightly smaller than the results quoted in
Ref.~\cite{BB96}. As both values are nearly equal, we shall use a
common value
\begin{equation}
a_2^\perp(\rho) = 0.15\pm 0.07 = a_2^\parallel(\rho)\,.
\end{equation}
The corresponding value of $\Delta$ is
$$\Delta_\rho^\perp = 1.24\pm 0.11 = \Delta_\rho^\parallel\,,$$
with a central value slightly larger than that used in
Ref.~\cite{BZ04}. The value of $a_2(K^*)$ has been determined in
Ref.~\cite{elena}. Again, we re-examine these sum rules for the present
paper. We find $a_2^\perp(K^*) = 0.11\pm 0.09$ and $a_2^\parallel(K^*) =
0.10\pm 0.08$, which is more
conveniently presented by
the difference between $a_2(K^*)$ and $a_2(\rho)$:
\begin{eqnarray*}
a_2^\perp(K^*) -a_2^\perp(\rho) & = & -0.04 \pm 0.02\,,\\
a_2^\parallel(K^*) -a_2^\parallel(\rho) & = & -0.03 \pm 0.02.
\end{eqnarray*}
As both differences are nearly equal, we shall use 
\begin{equation}
a_2(K^*) - a_2(\rho) = -0.04\pm 0.03
\end{equation}
for both polarisations. This translates into
$\Delta_{K^*}^\perp=1.18\pm 0.14$, with errors largely correlated with
those of $\Delta_\rho^\perp$. 

The value of $a_1(\rho)$ vanishes by G-parity.
The values of $a_1^{\parallel,\perp}(K^*)$ have been subject to some
controversy over the recent years, which was settled only very
recently; in this work, we use the values obtained in
Ref.~\cite{BZa1}: 
\begin{equation}
a_1^\perp(K^*) = 0.04\pm 0.03,\qquad a_1^\parallel(K^*) = 0.03\pm
0.02\,.
\end{equation}
All odd Gegenbauer moments, i.e.\ the antisymmetric contribution to
$\phi(u)$, can be resummed using the same power-like behaviour of
large moments as in the BT model. 
This model is also discussed in
Ref.~\cite{angi} and normalised to $a_1$; we include terms up to $n=9$.

Twist-3 and -4 DAs of vector mesons have been studied in
Refs.~\cite{BBKT,BB98_2}. The results are complete for 
mesons with definite G-parity
(with equal-mass quarks), but miss certain G-parity-breaking
corrections. A complete analysis of all these corrections is in
preparation \cite{prep}; here, we include those
results that are already available \cite{0609}. In Ref.~\cite{BBKT}, the
two-particle twist-3 DAs $g_\perp^{(v,a)}$ and $h_\parallel^{(s,t)}$
have been expressed in terms of integrals over the twist-2 DAs
$\phi_{\perp,\parallel}$ and the three-particle twist-3 DAs $\cal
A,V,T$. These integral relations are complete, but the explicit 
expressions for the three-particle twist-3 DAs given in \cite{BBKT} have to be
extended to include G-parity-breaking corrections as follows:
\begin{eqnarray}
{\cal A}(\underline{\alpha}) & = & 360 \alpha_1 \alpha_2 \alpha_3^2
\zeta_3^A \left\{ 1 + \lambda_3^A (\alpha_1 - \alpha_2) + \omega_3^A
  \left(\frac{7}{2} \alpha_3 - \frac{3}{2}\right)\right\},\nonumber\\
{\cal V}(\underline{\alpha}) & = & 360 \alpha_1 \alpha_2 \alpha_3^2
\left\{ \kappa_3^\parallel + \frac{3}{2}\,\zeta_3^A  \omega_3^V 
 (\alpha_1 - \alpha_2) + \kappa_3^\parallel \lambda_3^V  
\left(\frac{7}{2} \alpha_3 - \frac{3}{2}\right)\right\},\nonumber\\
{\cal T}(\underline{\alpha}) & = & 360 \alpha_1 \alpha_2 \alpha_3^2
\left\{ \kappa_3^\perp + \frac{3}{2}\,\zeta_3^A  \omega_3^T 
 (\alpha_1 - \alpha_2) + \kappa_3^\perp \lambda_3^T  
\left(\frac{7}{2} \alpha_3 - \frac{3}{2}\right)\right\}.
\end{eqnarray}
Here $\zeta_3^A$ and $\omega_3$ are G-parity conserving quantities,
whereas $\kappa_3$ and $\lambda_3$ are G-parity breaking.
As $\kappa_3^\perp$ contributes to the form factor only 
at ${\cal O}(m_V^2/m_b^2)$, and the $\lambda_3$ parameters are of
non-leading conformal spin, we set, in the
present analysis, the central values of 
all these parameters to zero and only take into account $\kappa_3^\parallel$.
A QCD sum rule estimate of this parameter yields \cite{prep,0609}
\begin{equation}
\kappa_3^\parallel(1\,{\rm GeV}) = 0.001\pm 0.001.
\end{equation}
The effect of non-zero values of $\kappa_3^\perp$ and $\lambda_3$ is
taken into account by the variation of these parameters around zero
within the range given in Table~\ref{tab:1}; the dependence of
$T_1^{K^*}$ on NLO G-parity breaking parameters is very small, as expected.

The two-particle twist-4 DAs $h_3$ and ${\mathbb A}_\perp$ 
have been discussed in
Ref.~\cite{BB98_2}; they are given by integrals over chiral-odd twist-4
three-particle DAs. The determination of the conformal-expansion coefficients 
of the latter is complicated by the fact that they contain ``kinetic''
mass-correction terms given by twist-2 matrix elements, which, 
to date, have not been obtained in a
closed form, but have to be unravelled order by
order in the conformal expansion. In addition, the direct
determination of the ``genuine'' twist-4 corrections from QCD sum
rules becomes
increasingly complicated at higher-order conformal spin. For that
reason, we invoke an alternative estimate of these corrections based on
the renormalon-model developed in Ref.~\cite{renormalon}. The general 
idea of this technique is to estimate matrix elements of ``genuine'' 
twist-4 operators by the quadratically divergent
contributions that appear when the matrix elements are defined using a
hard UV cut-off. In this way, three-particle twist-4 DAs
can be expressed in terms of the leading-twist DA $\phi_\perp$ 
\cite{renormalon}:
\begin{eqnarray}
T_1(\underline{\alpha}) = - T_3(\underline{\alpha}) & = & \zeta_4^T
\left[ \frac{\alpha_2\phi_\perp(\alpha_1)}{(1-\alpha_1)^2} -
  \frac{\alpha_1
    \phi_\perp(1-\alpha_2)}{(1-\alpha_2)^2}\right],\nonumber\\
T_2(\underline{\alpha}) = \phantom{-} T_4(\underline{\alpha}) & = &
-\frac{1}{2} \zeta_4^T
\left[ \frac{\phi_\perp(\alpha_1)}{1-\alpha_1} -
  \frac{\phi_\perp(1-\alpha_2)}{1-\alpha_2}\right],\nonumber\\
S(\underline{\alpha}) = - \tilde{S}(\underline{\alpha}) & = &
\frac{1}{2} \zeta_4^T
\left[ \frac{\phi_\perp(\alpha_1)}{1-\alpha_1} +
  \frac{\phi_\perp(1-\alpha_2)}{1-\alpha_2}\right].\label{ren}
\end{eqnarray}
The above formulas differ from those given in Ref.~\cite{renormalon}
by the change of argument $\alpha_2\to 1-\alpha_2$ in the second terms
on the right-hand side; this is to properly account for G-parity-breaking
effects \cite{lenz}. One prediction of the renormalon model is that
the two independent LO twist-4 couplings $\zeta_4^T$ and
$\tilde{\zeta}_4^T$ add up to 0, which is consistent with the direct
calculation from QCD sum rules \cite{BBK}. The above formulas also
allow one to estimate the ``genuine'' twist-4 G-parity breaking contributions 
to $T_i$ and $S$, $\tilde S$; we refrain from giving explicit formulas
in this paper, but refer to Ref.~\cite{prep}. For the calculation of the
contribution of twist-4 terms to $\xi$, we use two methods: firstly the
full renormalon model (\ref{ren}), and the corresponding
expression for ${\mathbb A}_\perp$ as given by the equations of motion
\cite{BB98_2} ($h_3=0$ in this model). This accounts for the genuine
twist-4 corrections; the ``kinetic''
corrections, as far as they are known, are added using truncated conformal
expansion. Secondly, we use truncated expansion for all twist-4 DAs,
describing G-parity-breaking terms by the values they assume in the
renormalon model, see Refs.~\cite{lenz,prep} for more details. The
predictions of both methods for the end-point behaviour of the DAs
near $u=0,1$ differ quite drastically; nonetheless, both prescriptions
given nearly the same result after integration
over $u$.

One more parameter that enters the kinetic mass corrections to twist-3
and -4 DAs, induced by the equations of motion, are the quark masses
$m_{s,u,d}$. We choose $\overline{m}_s(2\,{\rm GeV}) =
(0.10\pm 0.02)\,$GeV, which is in accordance with both lattice
\cite{lattms} and QCD sum rule calculations \cite{SRms}, and let
$m_{u,d}=0$.

\begin{figure}[tb]
$$\epsfsize=0.45\textwidth\epsffile{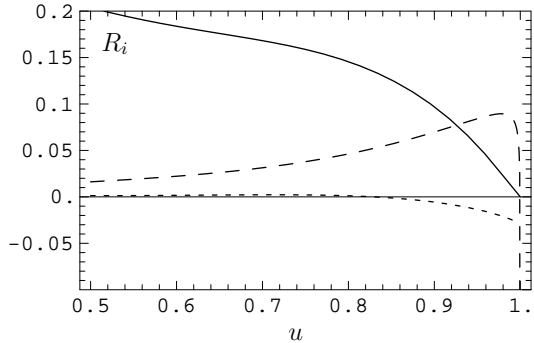}$$
\vspace*{-30pt}
\caption[]{Contribution of $R_i$ to the convolution integral in
  (\ref{SR}) as a
  function of $u$. Solid line: $f^\perp_\rho R_1(u)$, long dashes:
  $f^\parallel_\rho (m_\rho/m_b) R_2(u)$, short dashes: $f^\perp_\rho
  (m_\rho/m_b)^2  R_3(u)$.}\label{fig:0}
\end{figure}

With all DAs available, we can now assess the respective size of
the contributions of the various $R_i$ to the sum rule (\ref{SR}). To
this end, we plot, in Fig.~\ref{fig:0}, the functions $R_i$ for the
$\rho$ meson, multiplied by the corresponding weight factors, for
$u>0.5$ which is about the smallest value of $u_0$. The plot
clearly shows  that $R_1$ is dominant. It also shows that $R_{2,3}$
exhibit (integrable) end-point singularities for $u\to 1$. Based on
these results, we expect the impact of the first neglected 
term in the light-cone
expansion, which is ${\cal O}(m_V^3/m_b^3)$, to be very small.

\begin{figure}[p]
$$
\epsfxsize=0.45\textwidth\epsffile{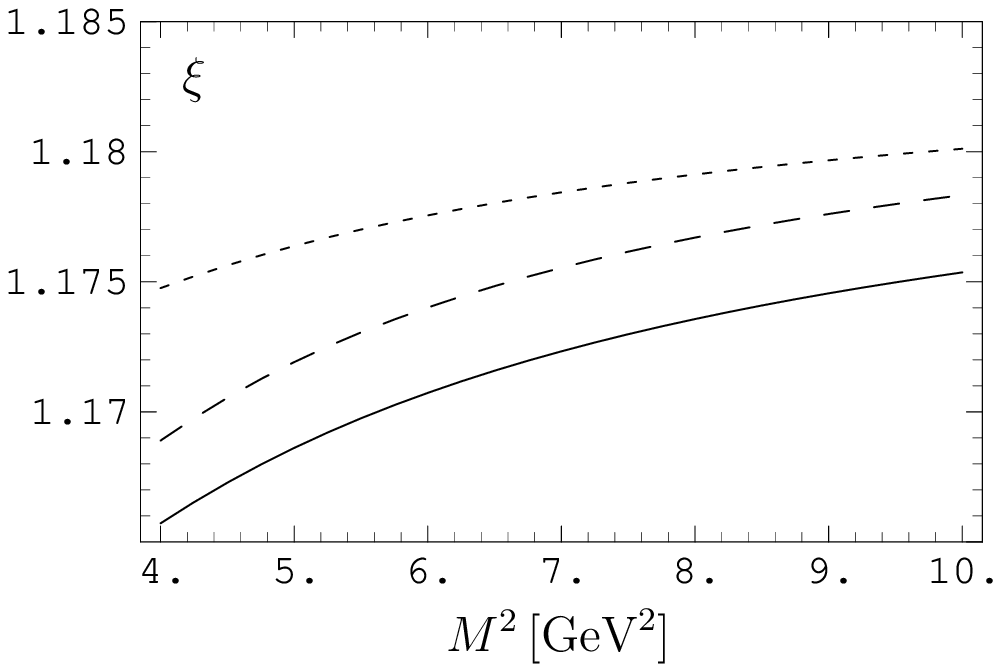}\qquad
\epsfxsize=0.45\textwidth\epsffile{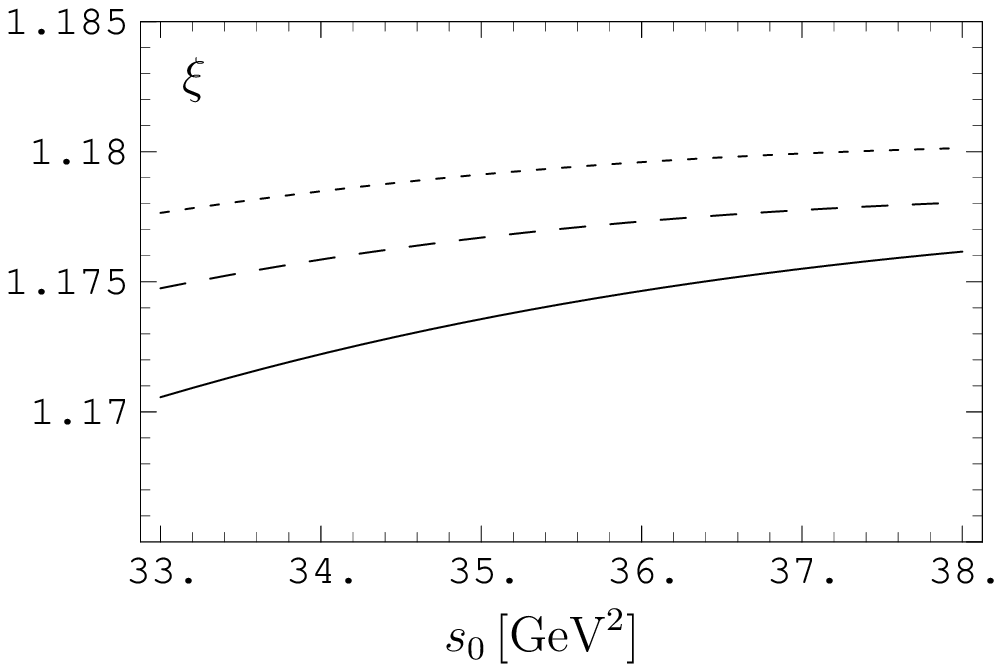}$$
\vspace*{-30pt}
\caption[]{Left panel: $\xi$ as a function of the Borel parameter $M^2$
  for $s_0 = 35\,{\rm GeV}^2$ and central values of the input
  parameters. Right panel: $\xi$ as a function of the continuum
  threshold $s_0$ 
  for $M^2 = 8\,{\rm GeV}^2$ and central values of the input
  parameters. Solid lines: DAs in conformal expansion; long dashes: BT
  model \cite{angi} for twist-2 DAs; short dashes: BT model for
  twist-2 DAs and 
renormalon model for twist-4 DAs \cite{renormalon}.}\label{fig:1}
$$
\epsfxsize=0.45\textwidth\epsffile{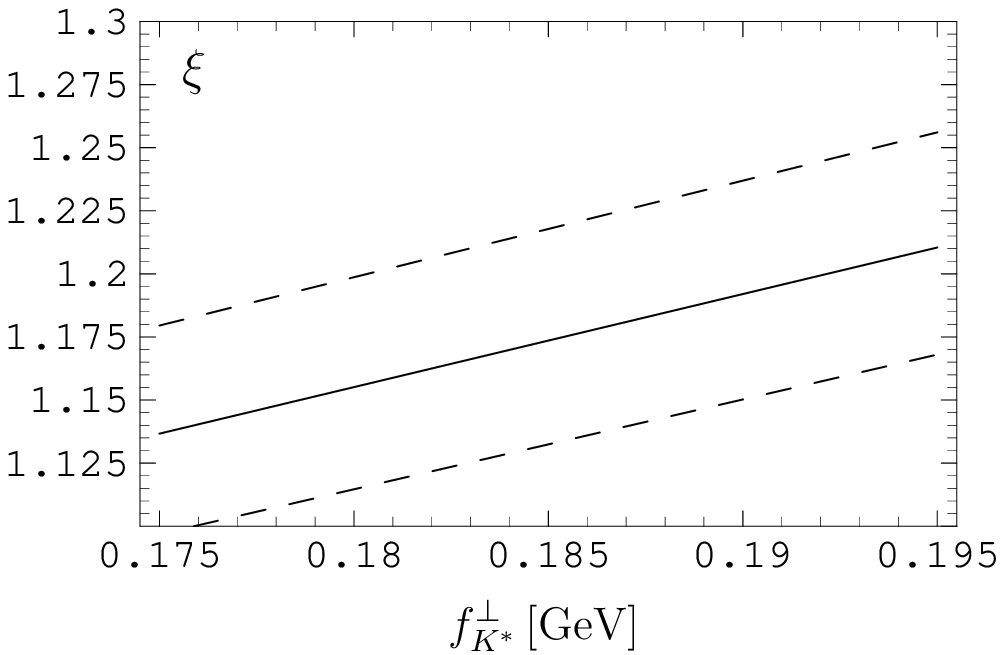}
$$
\vspace*{-30pt}
\caption[]{$\xi$ as a function of $f_{K^*}^\perp(1\,{\rm GeV})$. Solid
  line: $f_{\rho}^\perp(1\,{\rm GeV})=0.165\,{\rm GeV}$, dashed lines:
  $f_{\rho}^\perp$ shifted by $\pm 0.009\,$GeV.}\label{fig:2}
$$
\epsfxsize=0.45\textwidth\epsffile{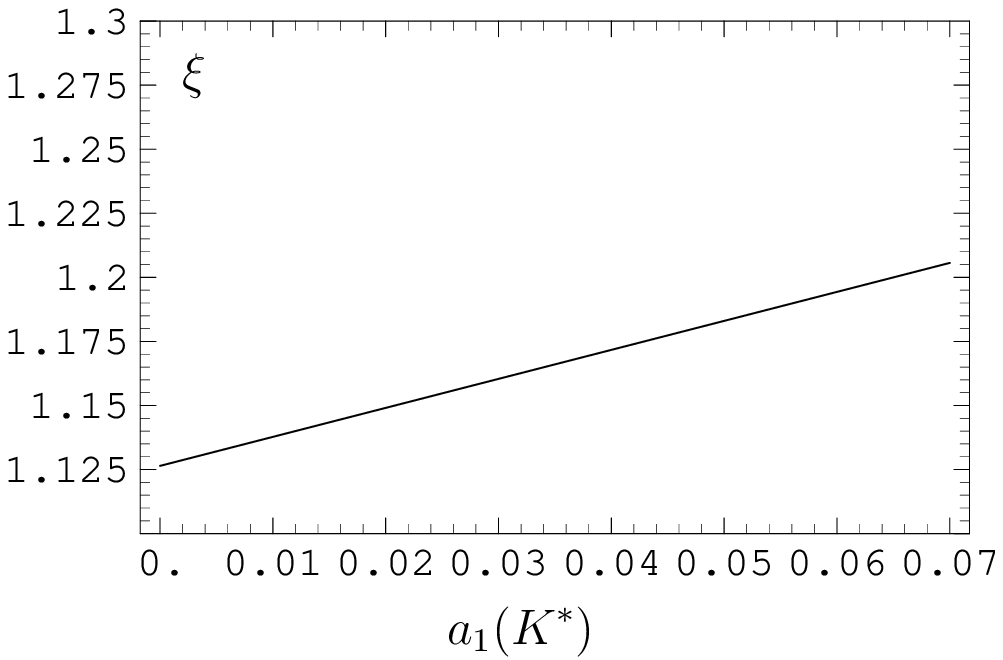}\qquad
\epsfxsize=0.45\textwidth\epsffile{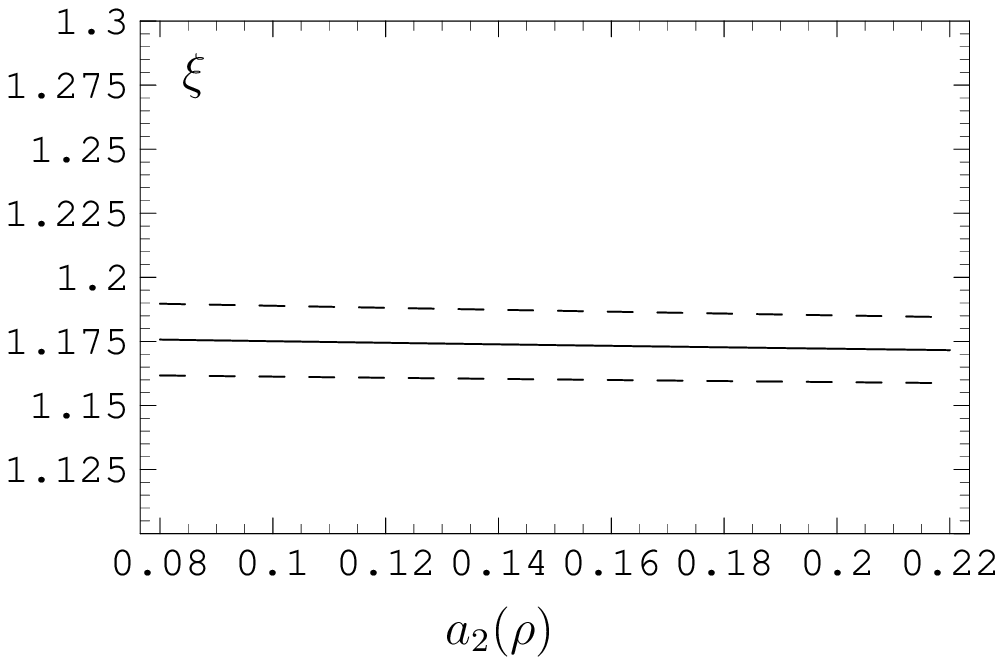}$$
\vspace*{-30pt}
\caption[]{Left panel: $\xi$ as a function of $a_1(K^*)$ at
  1~GeV. Right panel: $\xi$ as a function of $a_2(\rho)$ at 1~GeV. 
  Solid line: $a_2(K^*) = a_2(\rho)-0.04$; dashed lines: $a_2(K^*)$
  shifted by $\pm 0.02$. Longitudinal and transverse parameters
  $a_i^\parallel$ and $a_i^\perp$ are
  set equal.}\label{fig:3}
\end{figure}

Before we can evaluate the sum rule for $\xi$, we also have to discuss
the choice of $m_b$ and the sum-rule-specific parameters $M^2$ and
$s_0$. The good news is that, although numerator and denominator of
(\ref{xi}) both depend on $m_b$, $s_0$ and $M^2$, this dependence
cancels to a large extent in the ratio. The reason hereof is quite
evident from (\ref{SR}): $M^2$ controls the respective weights of
contributions of different $u$; as these contributions are nearly
equal in numerator and denominator of (\ref{xi}), except for
moderately sized SU(3) breaking, it follows that one can choose $M^2$
equal in $T_1^{B\to (\rho,K^*)}$ and  that the resulting dependence on
$M^2$ should be very small. This is borne out by the left panel of
Fig.~\ref{fig:1}, where we plot $\xi$ as function of $M^2$ for
central values of the input parameters and $s_0=35\,{\rm GeV}^2$. For
comparison, we also show $\xi$ calculated using the BT model
for the twist-2 DAs (long dashes) and, in addition, the renormalon-model
for twist-4 DAs (short dashes). All three calculations agree with one another
very well. The fact that the impact of the BT model is only minor
shows that the sum rules are sensitive only to a few gross
characteristics of the twist-2 DAs, but not to the details of their
behaviour near the end-point $u=1$. As for the renormalon-model DAs,
the difference to the truncated conformal expansion is most marked for 
small values of $M^2$,
which can be easily understood from the fact that for smaller $M^2$ the
weight of contributions from $u$ close to 1 gets enhanced
and hence the difference between the end-point behaviour of conformally
expanded DAs and renormalon-modelled DAs becomes more visible. The
right panel of Fig.~\ref{fig:1} illustrates the effect of a
variation of $s_0$ for fixed $M^2$, which is also small. The value of
$s_0$ sets the lower limit of the integral over $u$, and again the
dependence on $s_0$ largely cancels in the ratio as the integrands are equal
up to SU(3)-breaking effects. As $s_0$ itself is nearly independent of the
final-state meson, it is natural to choose the same value in both
numerator and denominator. As for $m_b$, it only enters in the ratio
$m_V/m_b$, which controls the respective contributions of $R_{1,2,3}$,
the lower limit of integration
$u_0=m_b^2/s_0$ and the Borel exponential $\exp(-m_b^2/(u
M^2))$. In the latter two parameters, a change of $m_b$ is effectively
compensated by a change of $s_0$ or $M^2$, which, as we have just
discussed, induces only very small variations of the sum-rule
result. The ratio $m_V/m_b$ changes from $0.185$ for $K^*$ and
$m_b=4.8\,$GeV to $0.193$ for $m_b = 4.6\,$GeV, which also has only 
very minor impact.
Based on these observations, we choose to evaluate $\xi$ for fixed $m_b =
4.8\,$GeV, 
$M^2=8\,{\rm GeV}^2$ and $s_0 = 35\,{\rm GeV}^2$ and attach to $\xi$ 
a corresponding uncertainty of $\pm 0.005$.

We are now in a position to obtain a result for $\xi$ and
estimate its uncertainty. The dominant uncertainty is due to the
dependence of $\xi$ on the chiral-odd twist-2 parameters. In
Fig.~\ref{fig:2} we plot $\xi$ as a function of
$f_{K^*}^\perp(1\,{\rm GeV})$, for various values of
$f_{\rho}^\perp(1\,{\rm GeV})$. The uncertainty in both parameters
causes an uncertainty in $\xi$ of $\pm 0.08$. In Fig.~\ref{fig:3},
left panel, we show the dependence of $\xi$ on $a_1(K^*)$, which
induces a change in $\xi$ by $\pm 0.03$; the variation of
$a_1^\perp(K^*)$ and $a_1^\parallel(K^*)$ as separate quantities
induces the same change. The right panel shows the dependence on $a_2$
which is rather mild and causes $\xi$ to change by $\pm 0.02$. The
variation of the remaining parameters within the limits specified in
Table~\ref{tab:1} causes another $\pm 0.02$ shift in $\xi$, so that we
arrive at the following result:
\begin{eqnarray}
\xi = \frac{T_1^{B\to K^*}(0)}{T_1^{B\to \rho}(0)} &=& 1.17 \pm
0.08(f^\perp_{\rho,K^*}) \pm 0.03(a_1)
\pm 0.02(a_2) \pm 0.02(\mbox{twist-3 and -4})\nonumber\\
&&{} \pm 0.01 (\mbox{sum-rule
  parameters, $m_b$ and twist-2 and -4 models})\nonumber\\
&= &1.17\pm 0.09\,.\label{resxi}
\end{eqnarray}
The total uncertainty of $\pm 0.09$ is
obtained by adding the individual terms in quadrature. Let us stress
again that the error of this result is dominated by far by parameter
uncertainties, and is nearly independent of the sum rule specific
parameters; it is also independent of $f_B$.

\section{\boldmath Determination of $|V_{td}/V_{ts}|$}\label{sec:3}

Let us now turn to the calculation of the ratio of branching ratios
and the determination of $|V_{td}/V_{ts}|$. The Belle collaboration
has measured the quantity 
$$
R_{\rm exp} \equiv 
\frac{\overline{\cal B}(B\to (\rho,\omega)\gamma)}{\overline{\cal B}(B\to
  K^*\gamma)}\,,
$$
where $\overline{\cal B}(B\to (\rho,\omega)\gamma)$ is defined as the
CP-average $\frac{1}{2}[{\cal B}(B\to (\rho,\omega)\gamma)+{\cal
  B}(\bar B\to (\bar\rho,\omega)\gamma)]$ of 
$${\cal B}(B\to (\rho,\omega)\gamma) = \frac{1}{2}\left\{ {\cal B}(B^+\to
\rho^+\gamma) + \frac{\tau_{B^+}}{\tau_{B^0}}\left[ {\cal B}(B^0\to
\rho^0\gamma) + {\cal B}(B^0\to\omega\gamma)\right]\right\},
$$
and $\overline{\cal B}(B\to K^*\gamma)$ is the isospin- and
  CP-averaged branching ratio of the $B\to K^*\gamma$ channels. 
In 2005, Belle  reported a $5.1\sigma$ measurement \cite{Belle},
\begin{equation}\label{RexBelle}
R_{\rm exp}^{\rm Belle}= 
0.032\pm 0.008({\rm stat}) \pm 0.002({\rm syst})\,,
\end{equation}
followed by a $5.2\sigma$ measurement by BaBar in 2006 \cite{BaBar2}:
\begin{equation}\label{RexBaBar}
R_{\rm exp}^{\rm BaBar} = 0.024  \pm 0.005\,,
\end{equation}
where the statistical and systematical uncertainty are added in quadrature.
HFAG combines both results into the average \cite{HFAG}
\begin{equation}\label{RexHFAG}
R_{\rm exp}^{\rm HFAG}= 0.028\pm 0.005\,.
\end{equation}

Within QCD factorisation, and using the notations of Ref.~\cite{BoBu},
the amplitude for $B\to V\gamma$ can be written as
$$A(\bar B\to V\gamma) = \frac{G_F}{\sqrt{2}}\left[ \lambda_u a_7^u(V\gamma) +
  \lambda_c a_7^c(V\gamma)\right] \langle V\gamma | Q_7 | \bar B\rangle\,,$$
where $\lambda_q$ are products of CKM matrix elements and the
  factorisation coefficients $a_7^{u,c}$ consist of Wilson
  coefficients and non-factorisable corrections from hard scattering
  and annihilation; explicit expressions can be found in
  Ref.~\cite{BoBu}. $a_7^{u,c}$ depends in particular on the form factor
  $T_1$ and the twist-2 DA $\phi_{V;\perp}$. The theoretical expression
  for $R$ is then given by
\begin{eqnarray}
R_{\rm th} &=& \left|\frac{V_{td}}{V_{ts}}\right|^2 \frac{1}{\xi^2}
  \left(\frac{1-m_\rho^2/m_B^2}{1-m_{K^*}^2/m_B^2}\right)^3 \left|
  \frac{a_7^c(\rho\gamma)}{a_7^c(K^*\gamma)}\right|^2 \left( 1 +
  {\rm Re}\,(\delta a_\pm + \delta a_0) \left[\frac{R_b^2 - R_b
  \cos\gamma}{1-2 R_b \cos\gamma + R_b^2}\right]\right.\nonumber\\
& & \left. + \frac{1}{2}\left( |\delta a_\pm|^2 + |\delta a_0|^2\right)
  \left\{ \frac{R_b^2}{1-2 R_b \cos\gamma + R_b^2}\right\} \right)\label{Rth}
\end{eqnarray}
with $\delta a_{0,\pm}=
a_7^u(\rho^{0,\pm}\gamma)/a_7^c(\rho^{0,\pm}\gamma)-1$. Here, $\gamma$ is
one angle of the CKM unitarity triangle and $R_b$ one of its sides:
$$
R_b =
\left(1-\frac{\lambda^2}{2}\right)\frac{1}{\lambda}
\left|\frac{V_{ub}}{V_{cb}}\right|.
$$
Equation~(\ref{Rth}) differs from the expression given in Ref.~\cite{BoBu}
by the terms in $|\delta a|^2$ which were neglected in that paper. 
It is obtained in the SM, assuming that $\bar{\cal B}(B^0\to
\rho^0\gamma)\equiv\bar{\cal B}(B^0\to\omega\gamma)$, which
indeed should be the case up to a small difference in the decay
constants, a tiny difference in phase space and 
the sign of the contribution of weak annihilation (WA) diagrams,
which is also  small numerically. Equation~(\ref{Rth}) 
is also valid in extensions of the SM where the
CKM matrix is still unitary and the 
$a_7$ do not carry a new weak phase, for instance Minimal Flavour
Violation; in this case new physics could change the values of
$\delta a_{0,\pm}$. 

\begin{figure}[tb]
$$\epsfxsize=0.45\textwidth\epsffile{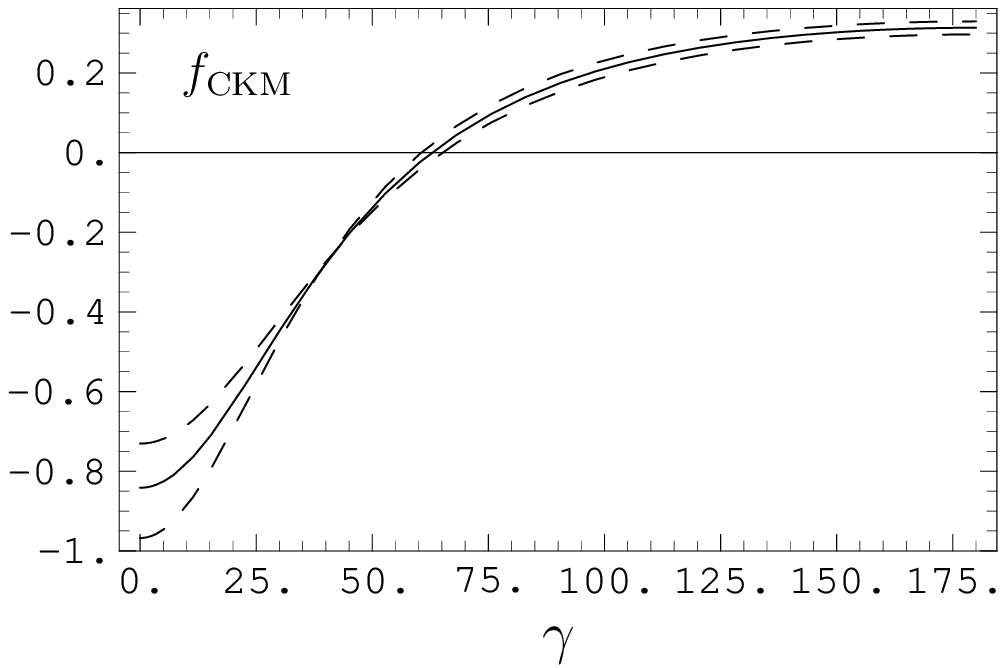}\qquad
\epsfxsize=0.45\textwidth\epsffile{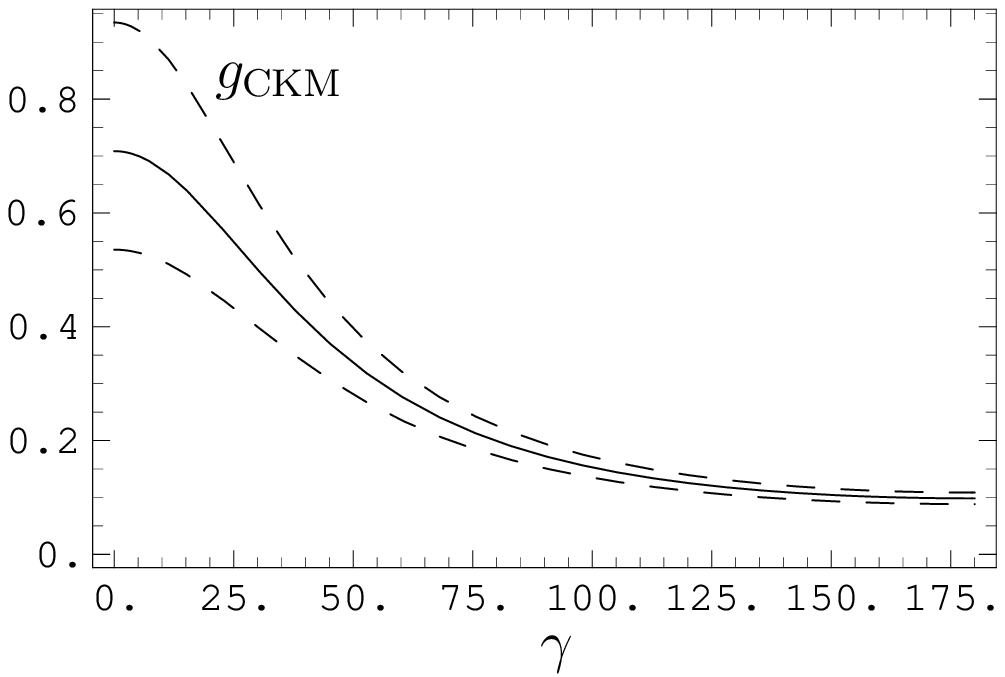}$$
\vspace*{-30pt}
\caption[]{Left panel: the CKM factor $f_{\rm CKM}$ 
in Eq.~(\ref{Rth}) (square brackets) as a
  function of $\gamma$ for $|V_{ub}/V_{cb}| = 0.106$ (solid line) and 
$|V_{ub}/V_{cb}| =  0.106\pm 0.008$ (dashed lines). Right panel: ditto
for $g_{\rm CKM}$ (curly brackets).}\label{fig:f}
\end{figure}

Let us first discuss the dependence of (\ref{Rth}) on CKM parameters,
described by the terms in square and curly brackets. The up-to-date value of 
$|V_{ub}/V_{cb}|$, as provided by the heavy flavour averaging group
HFAG in March 2006, is, adding errors in quadrature \cite{HFAG}:
$$
\left|\frac{V_{ub}}{V_{cb}}\right| = 0.106 \pm 0.008.
$$
In Fig.~\ref{fig:f} we plot the CKM factors
$$
f_{\rm CKM} = \frac{R_b^2-R_b\cos\gamma}{1-2 R_b\cos\gamma +
  R_b^2}\,,\qquad
g_{\rm CKM} = \frac{R_b^2}{1-2 R_b\cos\gamma + R_b^2}
$$
as functions of $\gamma$; the uncertainty induced by $R_b$ is
small. What is the currently preferred value of $\gamma$? HFAG is yet
to provide averages of the individual results obtained by BaBar and
Belle, so we use the value quoted 
by the UTfit collaboration in March 2006 \cite{UTfit}:
\begin{equation}
\label{gUTfit}
\gamma_{\rm UTfit} = (71\pm 16)^\circ\,,
\end{equation}
which is obtained from tree processes only and hence can be assumed to
be free of new physics. We then obtain
\begin{equation}
f_{\rm CKM} = 0.07\pm 0.12\,,\qquad g_{\rm CKM} = 0.23\pm 0.07\,.
\end{equation}
As $f_{\rm CKM}$ is rather small for the angle $\gamma_{\rm UTfit}$, 
Eq.~(\ref{gUTfit}),
the contribution of the corresponding non-factorisable contributions 
to (\ref{Rth}),
collected in Re\,$\delta a_{0,\pm}$, is heavily
suppressed.

The parameters
$|a_7^c(\rho\gamma)|$ and $|a_7^c(K^*\gamma)|$ are almost exactly
equal, so we set $|a_7^c(\rho\gamma)/$ $a_7^c(K^*\gamma)|=1$. Is that
value likely to be changed by non-factorisable corrections? One type of
such corrections, soft-gluon emission from charm loops, has been
calculated in Refs.~\cite{power2,0609}; it amounts to a contribution to
$|a_7^c|$ of ${\cal O}(1/(m_b m_c^2))$ and is small by
itself ($\sim 2\%$), and even smaller is its SU(3)-breaking that
changes $|a_7^c(\rho\gamma)/a_7^c(K^*\gamma)|$ by less than 1\%. 
Another source of
corrections comes from terms in
$(a_7^u(K^*\gamma)-a_7^c(K^*\gamma))/a_7^c(K^*\gamma)$, multiplied
by the CKM factor $\sim \lambda^2 f_{\rm CKM}$,  which is tiny indeed.
Hence, we do not see any obvious source of significant corrections to
$|a_7^c(\rho\gamma)/a_7^c(K^*\gamma)|=1$.

The value of 
$\delta a_{0,\pm}$ is calculated in QCD
factorisation, which is accurate to ${\cal O}(\alpha_s)$, 
but misses, in general,
terms that are suppressed by inverse powers of $m_{b,c}$. The most
important of these power-suppressed corrections is weak annihilation, 
which can actually  be calculated in QCD
factorisation, at least at tree level. WA is CKM suppressed in $B\to
K^*\gamma$ and mainly affects
$a_7^u(\rho^\pm\gamma)$, but it is small for
$a_7^u((\rho^0,\omega)\gamma)$ because of a
suppression by Wilson coefficients and the fact that WA is
proportional to the electric charge of the quark involved, namely the
$u$ quark for $a_7^u(\rho^\pm\gamma)$ and the $d$ quark for
$a_7^u((\rho^0,\omega)\gamma)$. Although WA is formally
power-suppressed, it gets 
enhanced in $a_7^u(\rho^\pm\gamma)$ by large Wilson coefficients
and the absence of ${\cal O}(\alpha_s)$ suppression that affects other
non-factorisable corrections. Numerically, WA is actually as
large as the leading (in $1/m_b$) non-factorisable terms. 
In view of the importance of this
contribution, we treat WA in two
different ways: firstly, by using the QCD-factorised expression given
in Ref.~\cite{BoBu}; and secondly, by using the results obtained from QCD
sum rules on the light-cone \cite{power1,emi,prep2}.

The WA contribution to the amplitude of e.g.\ $B^-\to\rho^-\gamma$ can be 
written in the following way:
$$
A(B^-\to\rho^-\gamma)_{\rm WA} = \frac{G_F}{\sqrt{2}}\,\lambda_u
\left( C_1 + \frac{1}{3}\,C_2\right) \langle \rho^-\gamma | (\bar d u)_{V-A}
(\bar u b)_{V-A}| B^-\rangle\,.$$
In naive factorisation, the matrix element on the r.h.s.\ can be
written as
\begin{eqnarray*}
\langle \rho^-\gamma | (\bar d u)_{V-A} (\bar u b)_{V-A}| B^-\rangle
& = & \langle \rho^-| (\bar d u)_{V-A}|0\rangle \langle\gamma | 
(\bar u b)_{V-A}| B^-\rangle\\
&&{} + \langle \rho^-\gamma | (\bar d
u)_{V-A}|0\rangle \langle 0 | (\bar u b)_{V-A}| B^-\rangle\,.
\end{eqnarray*}
The second term on the r.h.s.\ has been shown to vanish in the chiral
limit, see Ref.~\cite{contact} for more details, 
so we will focus on the first term. Corrections
to naive factorisation are of ${\cal O}(\alpha_s)$, which may also
relax the chiral suppression of the second term. Neglecting the latter, 
we have, in the notations of Ref.~\cite{emi}:\footnote{Equation
  (\ref{BV}) differs from the definition given in Ref.~\cite{emi} by
  an overall sign. The reason is that in \cite{emi} the covariant 
derivative $D_\mu = \partial_\mu -
i e A_\mu$ was used, corresponding to a negative value of $e$. In
order to keep $F_{V,A}$ positive, we change the sign of the definition.}
\begin{eqnarray}
\langle \rho^-(p)\gamma(q) | (\bar d u)_{V-A} (\bar u b)_{V-A}|
B^-(p_B)\rangle &=&
\sqrt{4\pi\alpha}\,\frac{m_\rho f_\rho^\parallel}{m_B} \epsilon_\mu^{(\rho)}
\left\{F_V \epsilon^{\mu\nu\rho\sigma} \epsilon_\nu^{(\gamma)}p_{B\rho}
  q_\sigma\right.\nonumber\\
&&\left. - i F_A [\epsilon^{\mu(\gamma)} (p_B\cdot q) - q^\mu
    (\epsilon^{(\gamma)}\cdot p_B)]\right\}.\label{BV}
\end{eqnarray}
The form factors $F_{A,V}$ can be calculated in QCD factorisation
themselves; both $F_{A,V}$ are then equal, and to LO accuracy one has
\begin{equation}\label{hard}
F^{\rm QCDF}_{\rm WA} \equiv F_{A,V}^{\rm QCDF} = \frac{Q_u f_B}{\lambda_B}
\end{equation}
with $Q_u = 2/3$ the electric charge of the $u$ quark and $\lambda_B$
the first inverse moment of the $B$-meson DA $\phi_B$:
$$
\int_0^1 d\xi \,\frac{\phi_B(\xi)}{\xi} = \frac{m_B}{\lambda_B}\,.
$$
Equation (\ref{hard}) 
agrees with the result  obtained in Ref.~\cite{BoBu} by
direct calculation of the WA diagram. Corrections are
either of ${\cal O}(\alpha_s)$, and have been calculated in Ref.~\cite{chris},
or they are suppressed by powers of $1/m_b$. The dominant source of the
latter comes from photon-emission from a soft $u$
quark and has been calculated in Refs.~\cite{power1,emi}, together
with the perturbative photon emission giving rise to (\ref{hard}). The
emission of photons from a soft quark line is
governed by the parameter $\chi$, the so-called magnetic
susceptibility of the quark condensate, $\langle 0 | \bar q
\sigma_{\alpha \beta} q | 0\rangle_F = \sqrt{4\pi\alpha}\,Q_q \chi
\langle \bar q q \rangle F_{\alpha\beta}$, which has been discussed in
detail in Refs.~\cite{BBK,kivel}, together with higher-twist DAs of the
photon. Its contribution is, in the heavy quark limit, suppressed by
one power of $1/m_b$ with respect to (\ref{hard}), but at finite quark
mass its size is set by the dimensionless parameter $\chi\langle
\bar u u\rangle/f_B \approx 0.2$, with $\chi(1\,{\rm GeV}) = (3.15\pm
0.3)\,{\rm GeV}^{-2}$ \cite{kivel}, which is not really small. In
calculating the WA contribution to $\delta a_\pm$, we will
use both expressions for $F_{V,A}$: $F^{\rm QCDF}_{\rm WA}$, 
Eq.~(\ref{hard}), and $F^{\rm QCDSR}_{\rm WA}$
from the QCD sum rule calculation, see Ref.~\cite{prep2} for details.

Let us first discuss $\delta a_0$, where WA is suppressed and can be
neglected. Its dependence 
on hadronic parameters is controlled by
the factor $f_B/(T_1^{B\to\rho} \lambda_B)$;
it also depends, to a lesser extent,
on $f^\perp_\rho$ and the twist-2 DA $\phi_{\rho;\perp}$. 
To estimate the uncertainty of Re$\,\delta a_0$ and $|\delta a_0|^2$, 
we set $f_B = (0.205\pm 0.025)\,{\rm GeV}$, which is an average of quenched and
unquenched lattice calculations \cite{fBlatt1,fBlatt2} and QCD sum
rule determinations \cite{fBSR}. We also use $T_1^\rho = 0.27\pm
0.03$ from light-cone sum rules,\footnote{This value, and in
  particular its error, is quoted from
  our previous paper in Ref.~\cite{BZ04}, but is in agreement with the
  evaluation of Eq.~(\ref{SR}).} and $\lambda_B(1\,{\rm GeV}) = 
(0.46\pm 0.11)\,$GeV, obtained in Ref.~\cite{lambdaB}. This value
supersedes the guesstimate 
$\lambda_B = (0.35\pm 0.15)\,$GeV \cite{BBNS} used in
previous calculations and agrees with the value $(0.46\pm 0.16)\,{\rm GeV}$
found in Ref.~\cite{offen}. We evaluate all spectator-interaction
contributions, that is those involving $\lambda_B$, at the scale
$\mu^2_h = m_B^2-m_b^2$, which is of order $\sim \Lambda_{\rm QCD} m_b$ as
advocated in Ref.~\cite{BoBu}, but by a factor 2 larger; this is
motivated, in part,
by the fact that the anomalous dimensions governing the
renormalisation-group running of the Wilson coefficients are
given for 5 flavours only in Ref.~\cite{SD} and hence should not be
used at scales as small as $(\Lambda_{\rm QCD} m_b)^{1/2}\sim 1.5\,$GeV. 
We then need to evolve
$\lambda_B$ from 1~GeV to $\mu_h$, which can be done using the
following evolution relation \cite{Lange}:
\begin{equation}
\lambda_B^{-1}(\mu) = \lambda_B^{-1}(\mu_0)\left\{ 1 +
\frac{\alpha_s}{3\pi}\,\ln \frac{\mu^2}{\mu_0^2}\left( 1 - 2
\sigma_B(\mu_0)\right)\right\},
\end{equation}
where $\sigma_B(1\,{\rm GeV}) = 1.4\pm 0.4$ is given by an integral
over the $B$-meson DA $\phi_B$ and was estimated in
Ref.~\cite{lambdaB}. We then have
$$
\lambda_B(\mu_h) = (0.51\pm 0.12)\,{\rm GeV}.
$$
We can now cast most of the
dependence of $\delta a_0$ on hadronic input parameters into a
dependence on $\lambda_B(\mu_h)$ only, varying it in the interval $\lambda_B =
(0.51^{+0.20}_{-0.11})\,$GeV. 
We also allow for 20\% power-suppressed corrections to the leading (in
$1/m_b$) non-factorisable corrections and find
\begin{eqnarray}
{\rm Re}\,\delta a_0 &=& 0.06\pm 0.02(\lambda_B,f_B,T_1)\pm 0.06({\cal
  O}(1/m_b))\,, \nonumber\\
|\delta a_0|^2 & = & 0.014\pm
  0.004(\lambda_B,f_B,T_1)^{+0.017}_{-0.009}({\cal
  O}(1/m_b))\,.\label{resa0}
\end{eqnarray}
Let us now turn to $\delta a_\pm$.
Neglecting the effect of WA, one has $\delta a_\pm =
\delta a_0$. 
Varying $\lambda_B$, $f_B$ and $T_1$ as before, and allowing for 20\%
power-suppressed corrections to leading non-factorisable contributions, we find
\begin{eqnarray}
{\rm Re}\,\delta a^{\rm QCDF}_\pm &=& 
-0.19\pm 0.09(\lambda_B,f_B,T_1)\pm 0.06({\cal
  O}(1/m_b))\,,\nonumber\\
|\delta a^{\rm QCDF}_\pm|^2 &=& 
\phantom{-}0.05^{+0.04}_{-0.03}(\lambda_B,f_B,T_1)^{+0.02}_{-0.01}({\cal
  O}(1/m_b))\,,\label{resapQCDF}
\end{eqnarray}
in QCD factorisation and
\begin{eqnarray}
{\rm Re}\,\delta a^{\rm QCDSR}_\pm &=& 
-0.06\pm 0.04({\rm SR})\pm 0.06({\cal
  O}(1/m_b))\,,\nonumber\\
|\delta a^{\rm QCDSR}_\pm|^2 &=& 
\phantom{-}0.02\pm 0.01({\rm SR})^{+0.03}_{-0.02}({\cal
  O}(1/m_b))\,,\label{resapQCDSR}
\end{eqnarray}
using QCD sum rules for the WA contribution. The SR error
reflects the dependence of the result on the QCD sum rule specific
parameters $M^2$ and $s_0$ and the value of $\chi$.

Taking everything together, we have
\begin{eqnarray}
R_{\rm th}^{\rm QCDF} &=& 
\left|\frac{V_{td}}{V_{ts}}\right|^2 \left[0.75\pm 0.11(\xi)\pm
0.03(a_7^{u,c},\gamma,R_b)\right],\nonumber\\
R_{\rm th}^{\rm QCDSR} &=& 
\left|\frac{V_{td}}{V_{ts}}\right|^2 \left[0.75\pm 0.11(\xi)\pm
0.02(a_7^{u,c},\gamma,R_b)\right].\label{resRth}
\end{eqnarray}
This result makes it clear that the theoretical uncertainty associated
with $\delta a$ is small and that the error is dominated by that of
$\xi$ --- the reduction of which is mostly a matter of more accurate
(lattice and QCD sum rule) calculations, but is not affected by
uncalculable $1/m_b$ corrections. 
Within the quoted accuracy, the two different methods
to calculate the WA contribution agree. We would like to stress here that
it is precisely the CKM suppression of $\delta a_{0,\pm}$ 
which also suppresses their uncertainties and 
renders the application of QCD factorisation to $B\to V\gamma$
viable.

We are now in a position to obtain values for $|V_{td}/V_{ts}|$. 
Comparing (\ref{resRth}) with the
experimental results (\ref{RexBelle}), (\ref{RexBaBar}) and
(\ref{RexHFAG}), we get  
\begin{eqnarray}
\left|\frac{V_{td}}{V_{ts}}\right|_{B\to V\gamma}^{\rm Belle} & =&  
0.207 \pm 0.016({\rm th}) \pm 0.027 ({\rm exp})  
\,,  \nonumber\\
\left|\frac{V_{td}}{V_{ts}}\right|_{B\to V\gamma}^{\rm BaBar} & = &  
0.179 \pm 0.014({\rm th})\pm 0.020 ({\rm exp}) 
 \,,\nonumber\\
\left|\frac{V_{td}}{V_{ts}}\right|_{B\to V\gamma}^{\rm HFAG} & =&  
0.192   \pm 0.014({\rm th})\pm 0.016 ({\rm exp})\,.\label{Rtrad} 
\end{eqnarray} 
These values can be compared with  that following from $R_b$,
$\gamma$ and the unitarity of the CKM matrix:
\begin{eqnarray}
\left|\frac{V_{td}}{V_{ts}}\right|_{\rm SM} & = &
\lambda (1+ R_b^2 - 2 R_b\cos\gamma)^{1/2}
= 0.216\pm 0.029\,.\label{RtSM}
\end{eqnarray}
Both results agree well within errors. As (\ref{RtSM}) is obtained from
tree-level processes only, it represents the ``true'' value of
$|V_{td}/V_{ts}|$ in the SM. 

A third determination of $|V_{td}/V_{ts}|$ can be obtained from $B$ mixing.
In the SM, we have
\begin{equation}\label{latt}
\frac{\Delta m_s}{\Delta m_d} = \frac{m_{B_s}}{m_{B_d}}\,
\frac{f_{B_s}^2 B_{B_s}}{f_{B_d}^2 B_{B_d}}
\left|\frac{V_{ts}}{V_{td}}\right|^2.
\end{equation}
The current world average for $\Delta m_d$ is
$\Delta m_d = (0.507\pm 0.005)\,{\rm ps}^{-1}$  \cite{PDG}.
$\Delta m_s$ has recently been measured by the CDF collaboration \cite{CDF},
\begin{equation}
\Delta m_s = 17.77\pm 0.10({\rm stat}) \pm 0.07({\rm syst})
{\rm ps}^{-1}\,,
\end{equation}
with an accuracy that exceeds 5$\sigma$ significance.
D0 provided a two-sided bound at 90\% CL\ \cite{D0}:
\begin{equation}
17\,{\rm ps}^{-1} < \Delta m_s < 21\,{\rm ps}^{-1}\,.
\end{equation}
The hadronic matrix elements in (\ref{latt}) are obtained from lattice
simulations. The most up-to-date results for the decay constants have
been obtained by the HPQCD group, using unquenched $n_f = (2+1)$ 
configurations \cite{HPQCD}:
$$
f_{B_s}/f_{B_d} = 1.20(3)(1)\,,
$$
where the first error is statistical and from chiral extrapolation and
the second comes from ``other uncertainties'' \cite{HPQCD}. The
particular strength of this calculation is that light quark masses as
small as $m_s/8$ could be reached, which implies that only a moderate
extrapolation to the physical chiral limit is required. As for the
ratio of $B_{B_{d,s}}$, the currently best result is obtained from
unquenched $n_f = 2$ calculations (JLQCD collaboration \cite{JLQCD}):
$$
B_{B_s}/B_{B_d}  = 1.017(16)(^{+56}_{-17})\,,
$$
where the first error is statistical and the second systematic. In
this calculation, the minimal light quark mass was $m_q = 0.5 m_s$,
which requires a more substantial  extrapolation to the physical
limit and is responsible for the large systematic
uncertainty. A combination of both results yields \cite{fBlatt2}:
\begin{equation} 
\frac{f_{B_s}B_{B_s}^{1/2}}{f_{B_d}B_{B_d}^{1/2}} =
1.210\left(^{+47}_{-35}\right), 
\end{equation}
where the errors have been added in quadrature. This procedure may be
problematic as it combines results with different systematic
effects, but yields the most reliable unquenched result to 
date.\footnote{A critical discussion of these lattice results, and their
impact on the constraints on new physics from $B$ mixing, can be found
in Ref.~\cite{rf}.} 
{}From this, one finds
\begin{equation}
\left|\frac{V_{td}}{V_{ts}}\right|_{\Delta m} = 
0.2060^{+0.0081}_{-0.0060}({\rm th})\pm0.0007({\rm exp})\,,
\end{equation}
which is the result obtained by the CDF collaboration \cite{CDF}.
This value, too, 
agrees with the two previous determinations.  Finally, one can compare
our result also to the results of global fits of the unitarity
triangle. The UTfit collaboration quotes, in September 2006, \cite{UTfit}
$$
\left|\frac{V_{td}}{V_{ts}}\right|_{\rm UTfit} = 0.202 \pm 0.008\,,
$$
whereas CKMfitter gets \cite{CKMfitter}
$$
\left|\frac{V_{td}}{V_{ts}}\right|_{\rm CKMfitter} = 0.201^{+0.008}_{-0.007}\,.
$$
Again, all values agree within errors.

\section{Summary and Conclusions}\label{sec:4}

In this paper we have presented a new analysis of the form-factor
ratio $\xi\equiv
T_1^{B\to K^*}/T_1^{B\to\rho}$ from QCD sum rules on the light-cone,
paying particular attention to the size of SU(3)-breaking effects. We
have obtained
$$\xi = 1.17\pm 0.09\,;$$
this value is nearly independent of QCD sum-rule-specific
parameters and the error is dominated by that of the tensor decay
constants $f_{\rho,K^*}^\perp$. A reduction of these errors by a
factor of two would reduce the total uncertainty to $\pm 0.06$. 
The numerical values of these
constants come mainly from QCD sum rules, partly from quenched lattice
calculations. A determination from unquenched lattice calculations
with reduced errors would be very desirable indeed.

We then have analysed the
non-factorisable corrections to $R\equiv\bar{\cal B}(B\to
(\rho,\omega)\gamma)/\bar {\cal B}(B\to K^*\gamma)$ in the framework of QCD
factorisation. The dominant power-suppressed correction comes from
weak annihilation diagrams that mostly affect $B^\pm\to
\rho^\pm\gamma$. We have estimated these corrections both in QCD
factorisation and using QCD sum rules, and find that the results agree
within errors; we will present a more detailed discussion of
power-suppressed corrections in a separate publication
\cite{prep2}. Our present best estimate of $R_{\rm th}$ is given in
Eq.~(\ref{resRth}). 
We then extracted the ratio of CKM matrix elements
$|V_{td}/V_{ts}|_{B\to V\gamma}$
from $R_{\rm exp}$ obtained by
BaBar and Belle, respectively, and averaged by HFAG, and 
find the values given in Eq.~(\ref{Rtrad}). 
Our  results for this parameter agree
well with all other  determinations available from various
sources as summarised in the previous section.
They also agree with the value extracted from $B$
mixing, using the new measurement of $\Delta m_s$ reported by the CDF
collaboration. Presently, there is no indication for new physics to be inferred
from these results.

\section*{Acknowledgments}
We would like to thank Sinead Ryan for a discussion of the current
status of lattice calculations of $B$ mixing parameters.

\section*{Addendum to v3}

Please note that in the arXiv version v2 of this paper, which  
is identical with the published version JHEP 04 (2006) 046,  
we used the BaBar bound  quoted in Ref.~\cite{BaBar}, $R_{\rm exp}^{\rm BaBar} 
< 0.029 \,\, {\rm
  at}\,\, 90\% \,\, {\rm CL}$,  
which was combined, by HFAG, with the Belle measurement to 
$R_{\rm exp}^{\rm HFAG}= 0.024\pm 0.006$ and resulted in $
\left| V_{td}/ V_{ts} \right|_{B\to V\gamma}^{\rm HFAG}  =  
0.179 \pm 0.014({\rm th}) \pm 0.022 ({\rm exp})$. These values have
changed with the BaBar measurement of $B(B\to (\rho,\omega)\gamma)$
reported in Ref.~\cite{BaBar2}; the corresponding 
new result for $|V_{td}/V_{ts}|$
is given in (29).

\appendix

\renewcommand{\theequation}{\Alph{section}.\arabic{equation}}
\renewcommand{\thetable}{\Alph{table}}
\setcounter{section}{1}
\setcounter{table}{0}

\section*{\boldmath Appendix: NLO Evolution of Twist-2 DAs}
\setcounter{equation}{0}

To leading-logarithmic accuracy, the (non-perturbative)
Gegenbauer moments 
$a_n$ in Eq.~(\ref{eq:confexp}) renormalize multiplicatively as
\begin{equation}
a^{\rm LO}_n(\mu^2) = L^{\gamma^{(0)}_n/(2\beta_0)}\, a_n(\mu_0^2),
\end{equation}
where $L = \alpha_s(\mu^2)/\alpha_s(\mu_0^2)$,
$\beta_0=(33-2N_f)/3$, and
the anomalous dimensions $\gamma^{(0)}_n$ are given by
\begin{eqnarray*}
\gamma^{\parallel(0)}_n &=&  8C_F \left(\psi(n+2) + \gamma_E - \frac{3}{4} -
  \frac{1}{2(n+1)(n+2)} \right),\\
\gamma^{\perp(0)}_n &=&  8C_F \left(\psi(n+2) + \gamma_E -
  \frac{3}{4} \right).
\end{eqnarray*}
To next-to-leading order accuracy, the scale dependence of the
Gegenbauer moments is more complicated and reads 
\cite{Mueller}
\begin{equation}
 a^{\rm NLO}_n(\mu^2) =  a_n(\mu_0^2) E_n^{\rm NLO} 
+\frac{\alpha_s(\mu^2)}{4\pi}\sum_{k=0}^{n-2} a_k(\mu_0^2)\,  
L^{\gamma_k^{(0)}/(2\beta_0)}\, d^{(1)}_{nk},  
\end{equation} 
where 
$$
E_n^{\rm NLO} =  L^{\gamma^{(0)}_n/(2\beta_0)}\left\{1+ 
            \frac{\gamma^{(1)}_n \beta_0 -\gamma_n^{(0)}\beta_1}{8\pi\beta_0^2}
                 \Big[\alpha_s(\mu^2)-\alpha_s(\mu_0^2)\Big]\right\}
$$
with $L=\alpha_s(\mu)/\alpha_s(\mu_0)$, $\beta_1 = 102-(38/3)N_f$; 
$\gamma^{(1)}_n$ are the diagonal two-loop anomalous dimensions, which
have been calculated, for the vector current, in Ref.~\cite{Floratos},
and, for the tensor current, in Ref.~\cite{Haya}.
The mixing coefficients $d^{(1)}_{nk}$, $k\le n-2$, 
are given in closed form in Ref.~\cite{Mueller}; these formulas are valid for
arbitrary currents upon substitution of the corresponding one-loop
anomalous dimension.\footnote{We thank D. Mueller for correspondence on 
this point.}

For the lowest moments $n=0,1,2$ one has, explicitly:
$$
\gamma_0^{\parallel(1)} =0\,, \qquad 
\gamma_1^{\parallel(1)} = \frac{23110}{243} - \frac{512}{81}\, N_f\,,
   \qquad  
\gamma_2^{\parallel(1)} =  \frac{34072}{243}-\frac{830}{81}\, N_f\,,
$$
\begin{equation}
\gamma_0^{\perp(1)} =\frac{724}{9} - \frac{104}{27}\,N_f\,, \qquad 
\gamma_1^{\perp(1)} = 124 - 8 N_f\,,\qquad  
\gamma_2^{\perp(1)} =  \frac{38044}{243}-\frac{904}{81}\, N_f\,,
\end{equation}  
and
\begin{eqnarray}
  d^{\parallel(1)}_{20} & = &
  \frac{35}{9}\,\frac{20-3\beta_0}{50-9\beta_0}
  \left(1-L^{50/(9\beta_0)-1}\right), \nonumber\\
  d^{\perp(1)}_{20} & = &
  \frac{28}{9}\,\frac{16-3\beta_0}{40-9\beta_0}
  \left(1-L^{40/(9\beta_0)-1}\right).
\end{eqnarray}

\end{document}